\documentstyle[11pt,newpasp,twoside,epsf]{article}
\def\edcomment#1{\iffalse\marginpar{\raggedright\sl#1\/}\else\relax\fi}
\reversemarginpar

\newbox\grsign \setbox\grsign=\hbox{$>$} \newdimen\grdimen \grdimen=\ht\grsign
\newbox\simlessbox \newbox\simgreatbox
\setbox\simgreatbox=\hbox{\raise.5ex\hbox{$>$}\llap
     {\lower.5ex\hbox{$\sim$}}}\ht1=\grdimen\dp1=0pt
\setbox\simlessbox=\hbox{\raise.5ex\hbox{$<$}\llap
     {\lower.5ex\hbox{$\sim$}}}\ht2=\grdimen\dp2=0pt

\def\chandra {{\it Chandra}}

\def\alpe {$\alpha_E$}

\def\La1215 {Ly$\alpha\lambda1215$}

\def\msun {${\rm M_{\bigodot}}$}

\newfont{\rura}{msam10 scaled \magstep1}

\def\gcm2	{{~g~cm$^2$}}

\def\scm2    {{~cm$^2$s$^{-1}$}}
\def\cm2    {{~cm$^2$}}

\def\doublespace {\smallskipamount=6pt plus2pt minus2pt
                  \medskipamount=12pt plus4pt minus4pt
                  \bigskipamount=24pt plus8pt minus8pt
                  \normalbaselineskip=24pt plus0pt minus0pt
                  \normallineskip=2pt
                  \normallineskiplimit=0pt
                  \jot=6pt
                  {\def\smallskip {\vskip\smallskipamount}}
                  {\def\medskip   {\vskip\medskipamount}}
                  {\def\bigskip   {\vskip\bigskipamount}}
                  {\setbox\strutbox=\hbox{\vrule 
                    height17.0pt depth7.0pt width 0pt}}
                  \parskip 12.0pt
                  \normalbaselines}
\def\onept5space {\smallskipamount=4pt plus2pt minus2pt
                  \medskipamount=8pt plus3pt minus3pt
                  \bigskipamount=16pt plus6pt minus6pt
                  \normalbaselineskip=16pt plus0pt minus0pt
                  \normallineskip=2pt
                  \normallineskiplimit=0pt
                  \jot=6pt
                  {\def\smallskip {\vskip\smallskipamount}}
                  {\def\medskip   {\vskip\medskipamount}}
                  {\def\bigskip   {\vskip\bigskipamount}}
                  {\setbox\strutbox=\hbox{\vrule 
                    height17.0pt depth7.0pt width 0pt}}
                  \parskip 12.0pt
                  \normalbaselines}
\def\myref#1  {\noindent \hangindent=24.0pt \hangafter=1 {#1} \par}
\def\bigref#1  {\noindent \hangindent=24.0pt \hangafter=2 {#1} \par}
\def\figs#1#2 {\item{#1} {#2} }

\begin{document}
\title{Spectral Energy Distributions of Quasars and AGN}
\author{Belinda J. Wilkes}
\affil{Smithsonian Astrophysical Observatory, 60 Garden St., Cambridge
MA 02138, USA}

\begin{abstract}
Active Galactic Nuclei (AGN) are multiwavelength emitters. To have any hope of 
understanding them, or even to determine their energy output,
we must observe them with many telescopes.
I will review what we have learned
from broad-band observations of relatively bright,
low-redshift AGN over the past $\sim 15$ years.
AGN can be found at all wavelengths but each provides a different
view of the intrinsic population, often with little overlap between
samples selected in different wavebands. 
I look forward to the
full view of the intrinsic population which we will obtain over the
next few years with surveys using today's new, sensitive
observatories. These surveys are already finding enough new and
different AGN candidates to pose the question ``What IS an AGN?".

\end{abstract}

\section{Introduction}
Unlike stars and galaxies, quasars and AGN are multi-wavelength emitters.
As a result obtaining a complete picture of an AGN is a challenging prospect
requiring observations with a wide variety of telescopes. Over the past
two decades, our multi-wavelength view of quasars and AGN has expanded
significantly thanks to the continuing increases in sensitivity 
(Sanders et al., 1989, Elvis et al. 1994, Haas
et al. 1998, Polletta et al. 2000, Haas et al. 2003).
The variety of the resulting
Spectral Energy Distributions (SEDs) grows with our
parameter space (Kuraszkiewicz et al. 2003)
and, while the contributing emission and absorption
mechanisms are well accepted, their 
relative importance, particularly as a function of AGN class,
remains a subject of debate.

Also hotly debated are the importance of orientation and absorption
in determining the SED, 
the relations between the various classes and the details of
the Unification picture. While
surveys at many wavelengths can efficiently find AGN, these surveys
provide different views of the AGN population, always selecting those
brightest in a particular waveband. It can be argued that
some wavebands provide a less biased view than others, (e.g. X-rays are
less affected by absorption, far-IR is independent of orientation).
However it is only by combining surveys in
different wavelength regions that
we can gain a view of the intrinsic population.
Only then can we hope to answer the many open questions that remain.

With the advent of new, sensitive observing facilities at many
wavelengths (X-RAY: Chandra, XMM-Newton, IR: 2MASS, SIRTF 
OPTICAL: 8-m telescopes, SDSS), multi-wavelength observations
are now possible for a significant fraction of the AGN population. 
Deeper, multi-wavelength surveys are finding
possible new varieties of AGN 
which raise the fundamental question: ``What is an AGN?".
Examples include the numerous, low-redshift red AGN found by 2MASS 
(Cutri et al., 2001)
and otherwise uninteresting, X-ray loud galaxies visible with Chandra
(Brandt et al. 2001).
Whether these new sources are truly AGN, how they relate to 
``traditional" AGN and how large a fraction of the population 
they constitute, are major open questions which will be addressed via
multi-wavelength follow-up.

Although AGN can be found in many wavebands, definitive classification is
challenging without optical/IR spectroscopic data. This is particularly true
given the lack of correspondence between the traditional optical class and
other characteristics in increasing subsets of the population
(e.g. IR/optical emission lines and X-ray flux (Genzel \& Cesarsky 2000), or
optical class and X-ray absorption (Wilkes et al. 2002)). 
The SDSS is
providing an unprecedentedly large sample to relatively faint 
optical flux limits, the means to classify AGN
with a wide range of SEDs. Similarly, 
SIRTF, successfully
launched in August 2003, will fill the last major gap in
our multi-wavelength picture of AGN reaching beyond the few bright and/or
nearby sources, to the bulk of the population in the mid- and far-IR
for the first time. 

With this unprecedented combination of powerful, multi-wavelength
observatories and the many planned and in progress surveys being
carried out (GOODS: Giavalisco et al. 2003,
ChaMP: Kim et al, 2003a, SWIRE: Lonsdale et al, 2003),
we are poised
to take great strides in our understanding of the intrinsic population 
of AGN, and their structure and evolution, as well as
the larger question of the importance of accretion
to the energy budget of the universe.

In this article, I review the observational components of AGN SEDs
along with
the physical structure and emission mechanisms believed to contribute 
in the various wavebands.
The shape and variety of the SEDs, colors, flux ratios and other 
properties are discussed as a function of AGN properties and in
comparison with models.
I conclude with the prospects for answering the question: ``What is an AGN?"
and thus of viewing the intrinsic population and constraining AGN models.

\section{The AGN Spectral Energy Distribution Observed}
\subsection{General Characteristics}
 

\begin{figure}
\plotfiddle{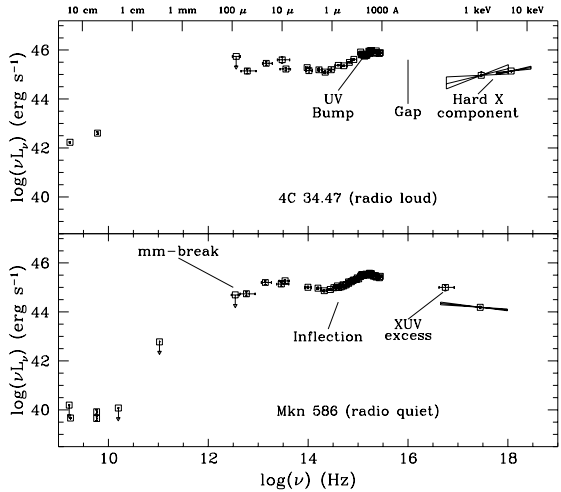}{3in}{-90}{35}{35}{-140}{260}
\vspace*{-0.9in}
\caption{Radio $--$ X-ray spectral energy distribution of
radio-loud (upper) and radio-quiet (lower) quasar, Elvis et al. (1994).}
\label{SED}
\end{figure}

Figure~\ref{SED} shows examples of fairly well-observed SEDs for both
a radio-quiet (RQQ) and a radio-loud (RLQ)
quasar. Both classes show peaks in their
energy output in the infra-red (IR bump)  and optical (``Big Blue Bump")
wavebands.
The IR bump is generally attributed to
thermal emission from dust at a wide range of temperatures, $\sim
50-$1000 K and the Big Blue Bump in thermal emission
from the gas in an accretion disk.
The relative strengths of the IR and Big Blue bumps varies but they are
generally comparable. The inflection between the two peaks, at
$\sim 1.5 \mu$ m, is due to the maximum dust temperature of
$\sim 2000$K caused by sublimation (Sanders et al. 1989).
In the X-ray region, $\sim 50$\% of both RLQs and RQQs have a soft X-ray excess
component thought to be the high energy tail of the Big Blue Bump.

At harder X-ray energies, power law
emission has differing slopes and relative strengths.
RQQs typically have slope, \alpe $\sim 1.0\pm0.5$ (where
F$_{\nu} \propto \nu^{-\alpha_E}$,
the range indicates a real spread in the observed slopes), while in RLQs
the slope is flatter ($\sim 0.5\pm0.5$) and the relative normalisation
about $\times 3$ higher (Wilkes \& Elvis 1987, Reeves \& Turner 2000).
The emission mechanisms are 
different: comptonisation of EUV photons in the Big Blue Bump for
RQQs (Gondek et al. 1996)
and synchrotron self-Compton scattering
of the radio photons in RLQs.
In lower luminosity AGN
reflected and/or scattered emission from cold/hot material surrounding
the X-ray source, such as a corona around the AD or the inner edge of
the dusty torus/disk (Mushotzky et al. 1993, 
Turner et al, 1997, Nandra et al 1997, Pounds et al. 2001), often dominate
the underlying power law. Strong Fe K$\alpha$ emission, originating
in cold and/or hot material, is present in many low luminosity AGN but
weaker/absent at higher luminosities
(Reeves \& Turner 2000).
Please see Brandt (this volume) for a more detailed review of
the X-ray emission of AGN.

\begin{figure}
\plotfiddle{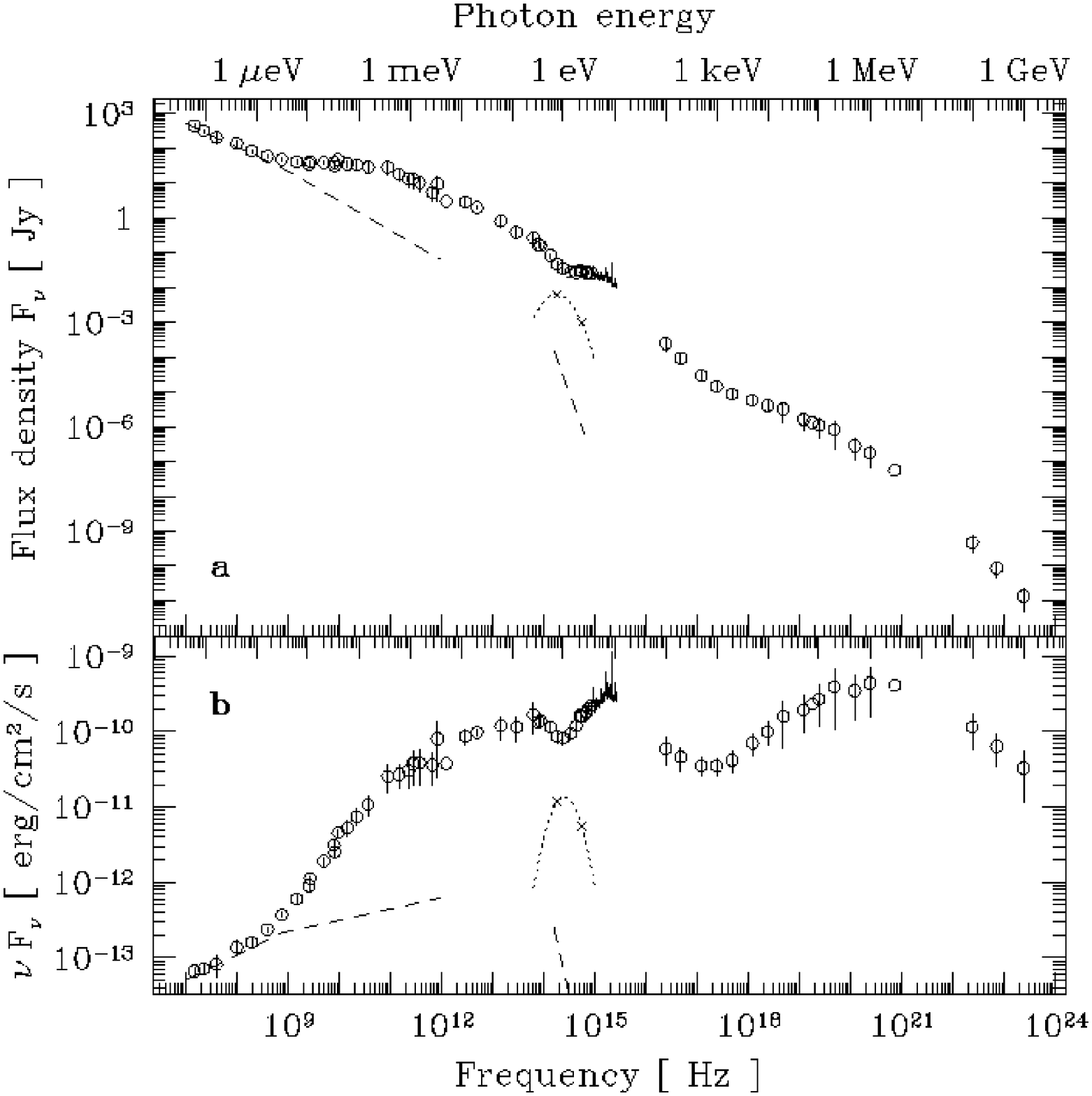}{3in}{0}{30}{30}{-100}{20}
\vspace*{-0.2in}
\caption{The radio$-\gamma$-ray SED of 3C273 on (a) F$_{\nu}$
and (b) $\nu$F$_{\nu}$ scale (Tuerler et al. 1999) showing the
smooth, radio$-$IR continuum emission 
typical of core-dominated RLQs.}
\label{3c273}
\end{figure}

The most notable difference between RLQs and RQQs is in the
radio waveband. In RQQs the SED turns over sharply in the far-IR/mm 
and radio emission is $\sim 100-1000 \times$
weaker than in RLQs. In RLQs the IR$-$radio continuum is smooth
with non-thermal emission contributing in both wavebands
(Figure~\ref{3c273}).

The far-IR cut-off in RQQs is well-determined in only a small number of
nearby sources which are bright in the far-IR. Constraints on its slope
are frequently steeper than the
$\nu^{2.5}$ characteristic of homogeneous synchrotron self-absorption.
Instead the far-IR emission is identified 
as grey-body emission from cool dust
(Chini et al. 1989, Hughes et al. 1993)

By contrast, the far-IR emission
from core-dominated RLQs smoothly extends into the radio, implying
a significant/dominant non-thermal component in both wavebands
(Figure~\ref{3c273}).
3C273 exhibits the correlated variability
characteristic of blazars (core-dominated RLQs viewed pole-on),
but even here the lack of variability in the
hottest part of the IR continuum indicates the presence of hot dust
(Tuerler et al. 1999).
ESA's Infrared Space Observatory (ISO), the most
sensitive IR satellite to date (prior to SIRTF, launched earlier this year),
facilitated observation of a larger number of quasars and AGN than in
the past. Comparison of the IR continua of RLQs and RQQs suggests that 
non-thermal IR emission dominates pole-on RLQs but decreases in
strength as the viewing angle increases so that thermal emission also
contributes in lobe-dominates RLQs (Haas et al. 1998, Polletta et al.
2000).

\begin{figure}
\plotfiddle{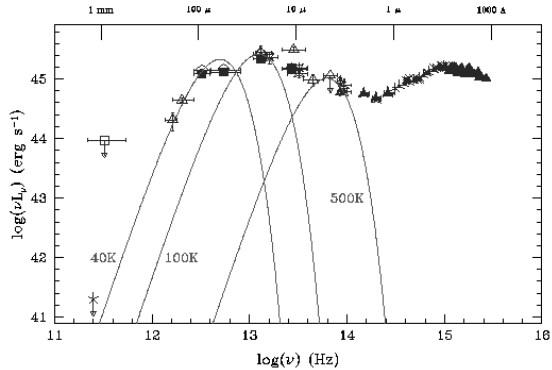}{3in}{0}{40}{40}{-120}{-0}
\vspace*{-1.2in}
\caption{mm$-$optical spectral energy distribution of PG1351+640
showing the wide range of temperature required to explain the full IR
emission as thermal emission from dust.
The curves show grey bodies at the marked temperatures
normalised to ``fit" the data.}
\label{irsed}
\vspace*{-0.2in}
\end{figure}

While both sources in Figure~\ref{SED} are typical of
the SEDs of broad-lined quasars in their class, the
relative strengths of the various components range by $\sim$
an order-of-magnitude from source to source,
even within the X-ray-bright subset of the population (Elvis et al. 1994).
This variety increases with less biased selection techniques such as the
X-ray (Kurasziewicz et al. 2003).

In the past 9 years since Figure~\ref{SED} was published
many more SEDs, mainly of low-redshift AGN, have been observed. SEDs
have many more data points, particularly in the IR and submm region
where ISO, IRAM and the JCMT have made significant steps forward.
However the general picture of the SEDs of ``normal" AGN has not changed
significantly. What has changed much more is our concept of what IS an AGN.

\subsection{Origin of the various components}

The physical picture for the origin of these components is
based on Unification of the 
AGN classes (Miller \& Antonucci 1983, Barthel 1989). The quasar's central
super-massive ($10^{7-9}$ \msun) black hole is surrounded by hot accreting
gas, primarily confined to an accretion disk (AD), emitting optical and UV
radiation to provide the Big Blue Bump. The IR bump originates in dust
with a wide range of temperatures, the hottest being directly
associated with the AGN while the cooler dust originates in the host
galaxy, again with a disk-like/torus geometry. X-ray 
and radio emission from the core varies on short timescales
($\sim 1-100$ lt. days) and so originates close to the
central black hole, interior to the AD. Some AGN ($\sim 10$\%)
have extended radio structure from jets and lobes on much larger
scales which will not be discussed here.

In a simple unification scenario broad-lined (Type 1) AGN are viewed face-on
while in narrow-lined (Type 2) AGN the broad emission line region (BELR)
the soft X-rays and much of the optical/UV emission from the AD
are hidden by the dust. The SEDs of type 2 AGN have
less prominent Big Blue Bumps and strong soft X-ray absorption 
than their type 1 counterparts. Comparisons between type 1 and 2
AGN at low redshift (Seyfert 1 and 2), where most
known Sy2s reside, agree with these expectations.
Their relatively low luminosities result in their optical
continua including a significant contribution from the host galaxy 
within which the central black hole and AGN reside.


To explain the broad IR continuum (Figure~\ref{irsed})
using pure thermal emission, the dust must have
a wide range of temperatures ($\sim 50-1000$K). Based upon
the presence of extended dust emission in nearby AGN (e.g. Cen A,
Genzel \& Cesarsky 2000), strong correlations between hot mid-IR
emission and the presence of an AGN (Heisler \& de Robertis 1999)
and weaker correlations between far-IR
emission and other AGN indicators
(Andreani et al. 2003), the IR continuum has been modeled using
2/3 components. A hot, obscuring dusty torus in an AGN (Pier \&
Krolik 1992, Granato \& Danese 1994) produces a narrow continuum
feature which is combined with a cooler, starburst component (Efstathiou
\& Rowan-Robinson 1995,
Rowan-Robinson 2000). Cool dust in the host galaxy may
also contribute.

\section{How well does the picture fit?}

\subsection{The Optical/UV Big Blue Bump}
The Big Blue Bump can be explained as thermal emission from gas in an accretion
disk (AD) with a wide range of temperatures.
ADs have a large number of parameters and can produce the
variety of observed SED shapes (Siemiginowska et al. 1995). Individual
fits which include the soft X-ray
excess require scattering of the AD photons in a lower density, hot
corona (Zycki et al. 1995, Laor et al. 1997).
As noted by Blaes (this volume), to
further constrain accretion disk models, an observational relation
between optical/UV colors and the black hole mass are required. Mass
measurements are available for a small number of low-redshift sources,
although new methods have been developed 
to estimate the mass at higher redshift (Vestergaard, this volume). 

As popular as accretion disks are, they are not unique in being
able to explain the optical$-$soft X-ray continuum of AGN.
Models invoking mildly optically thick cloud distributions 
(Czerny \& Dumont 1998, Collin-Souffrin et al. 1996) 
provide an alternative.

\subsection{IR continuum}

\begin{figure}
\plotfiddle{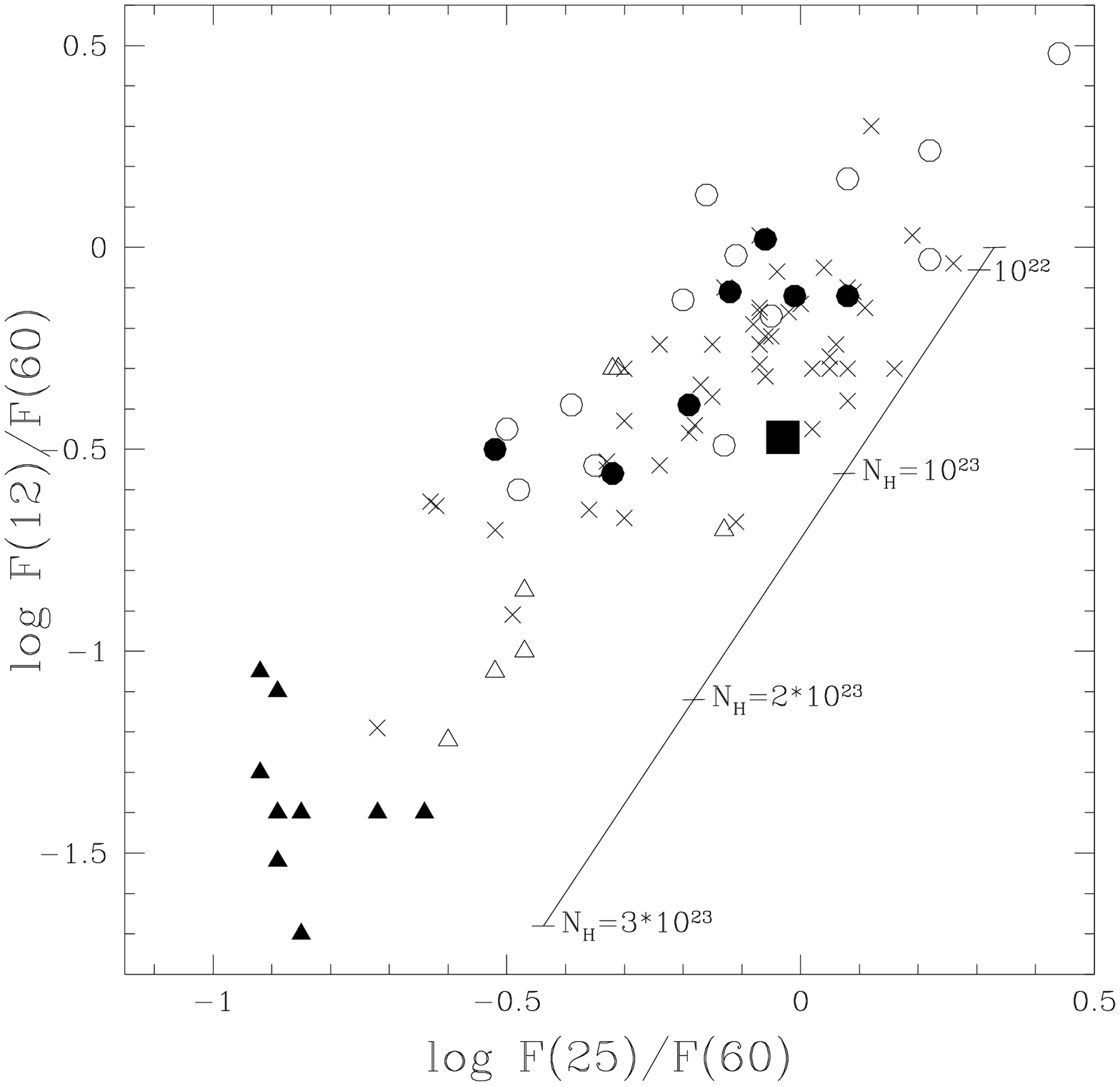}{3.5in}{0}{30}{30}{-180}{60}
\plotfiddle{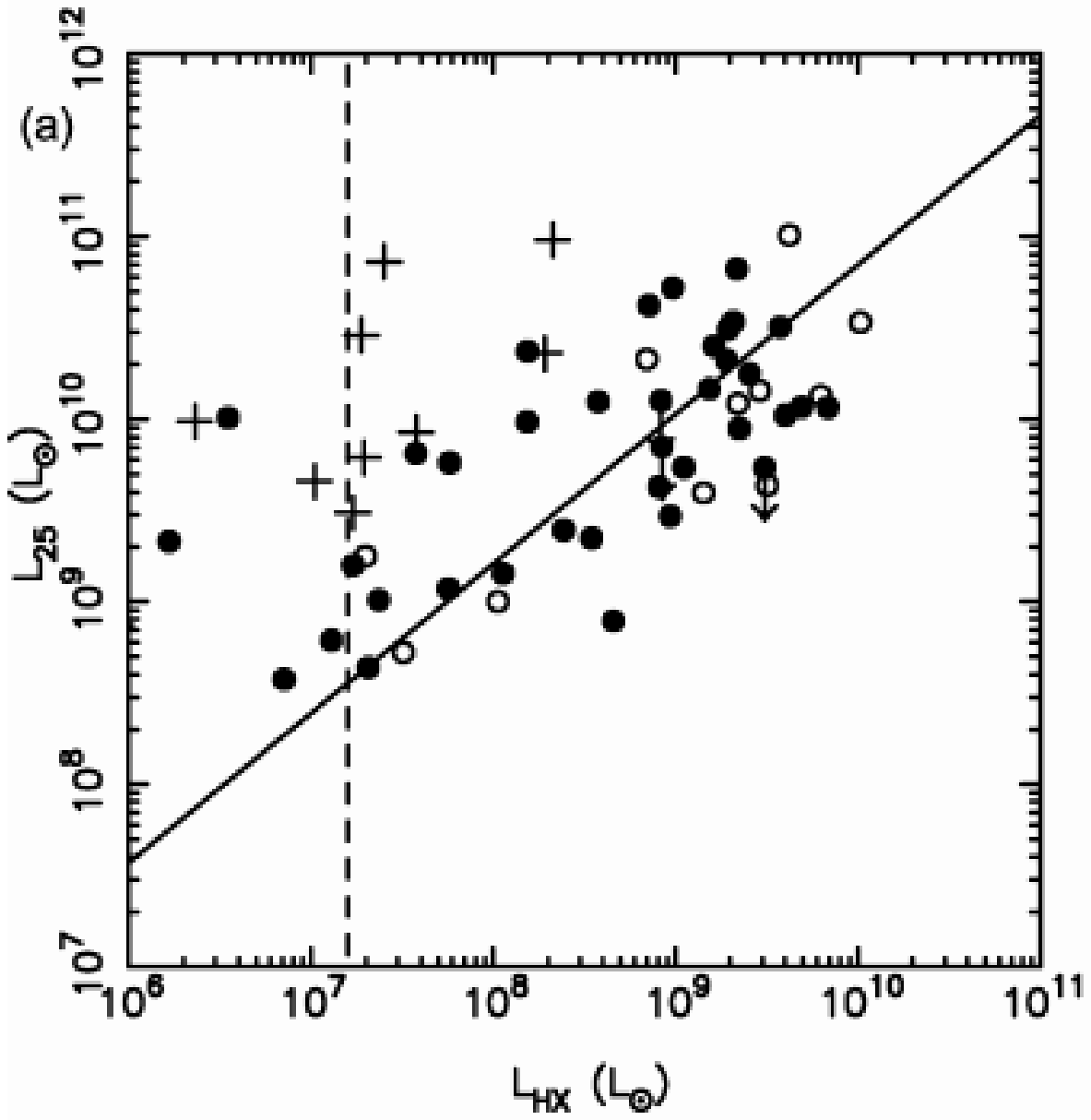}{3.5in}{0}{45}{45}{-30}{270}
\vspace*{-5.0in}
\caption{Left: Mid$-$IR colors for quasars compared with Seyfert 2 galaxies
(Heisler et al. 1997) showing that redenning of up to 
log N$_H \sim 23$ can explain the progression (Kuraszkiewicz et al.
2003, Fig.11). Right: The effects of host galaxy emission on the mid-IR colors of low
luminosity AGN (Lumsden \& Alexander 2001).}
\label{IRcolors}
\end{figure}

Spectroscopic observations from ISO (Clavel et al. 1998) show
a stronger near-mid IR continuum in Sy1 than in Sy2
galaxies causing a lower observed equivalent width 7.7 $\mu$m feature in
Sy1s. Orientation dependent near-IR emission is also clear
from comparison of Sy2s with and without hidden broad
line regions (Heisler et al. 1997). Further comparison with
Sy1s and quasars (Figure~\ref{IRcolors}, Kuraszkiewicz et al. 2003)
indicate obscuring column densities, log ${\rm N_H}
\sim 23$, as opposed to $\sim 24$ in earlier torus models (Pier \&
Krolik 1992). Lower column density (Kuraszkiewicz et al. 2003)
or clumpy (Zier \& Biermann 2002, Nenkova et al. 2002)
models for the obscuring material reproduce
the full range of temperatures in AGN IR continua with orientation
primarily responsible for the range in IR SED shapes.
This removes the need for a starburst-related component (Kuraszkiewicz et al. 2003),
although a two-component model is required in some 
low-redshift AGN (Genzel \& Cesarsky 2000).

An alternative suggestion 
invokes evolution to explain at least part of the range IR continuum shapes
(Haas et al. 2003). In this scenario (Sanders et al. 1988)
IR galaxies are young quasars/AGN where the
central engine is obscured by dust. The SED then changes from a young
state where the far-IR continuum dominates, through an AGN state in
which the
IR continuum is hotter and the Big Blue Bump appears and on to a dead
quasar where the latter component disappears. Although such a model
can, qualitatively, explain the range of SEDs observed in the PG sample
(Haas et al. 2003), there is no definitive evidence connecting SED shape
to the evolution/age of a source. 

\subsection{Orientation, obscuration and AGN Class}
\begin{figure}
\plotfiddle{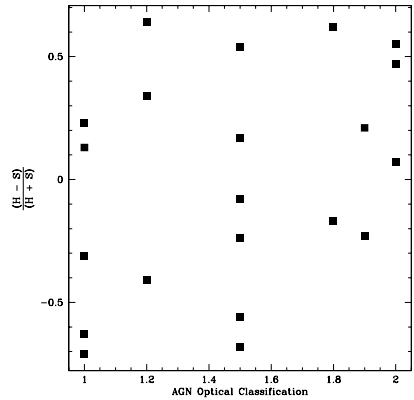}{3.5in}{0}{30}{30}{-100}{70}
\vspace*{-1.7in}
\caption{X-ray hardness ratios determined by the \chandra\
for 2MASS red AGN (Wilkes et al. 2002), as a function of optical
class demonstrating the lack of a relation between the two in this
sample.}
\label{HR_class}
\end{figure}

The unification scenario that type 2 AGNs are viewed through optically thick
material which hides the BELR and the central continuum source
is supported statistically by the general tendency for type 2 AGN to have absorbed X-ray spectra
(Turner et al. 1997, Awaki et al. 1997), a redder optical continuum, and stronger
galactic spectral features. Individual sources are
less straight-forward. Optical dust reddening is generally
lower than the equivalent X-ray gas absorption (Maiolino et al. 2001,
Risaliti et al. 2001). A few sources have been seen to change type, with
broad emission lines
appearing/disappearing and/or X-ray absorption varying (Matt et al. 2003).
Some type 1 or intermediate
sources have strong X-ray absorption (Page et al. 2001)
and there is little relation
between X-ray hardness ratio and AGN class (Wilkes et al. 2002,
Figure~\ref{HR_class}).
These results imply a more complex obscuring medium than the 
optically thick torus originally suggested. Current scenarios
generally involve 
high velocity, accelerating winds originating in a disk
(Konigl \& Kartje 1994, Murray \& Chiang 1995, Konigl this volume, Elvis this
volume). These models can also explain the high excitation, broad-absorption lines
visible in $\sim 10$\% of type 1 AGN. Variations and
differing lines-of-sight to continuum and line regions 
due to clumpiness and variation in the wind itself
can explain the variety of properties seen in individual sources. However to
provide real constraints on the structure and geometry of the absorbing
material, we need multi-wavelength observations of 
the intermediate sources discussed above.

\subsection{The Effects of Luminosity}
Sy2s, which are lower luminosity and more obscured
than Sy1s, often show 
optical/UV host galaxy spectral features. 
Galaxy light peaks in the wavelength region close to the minimum between
the IR and OUV bumps in AGN so the strength of this inflection
depends on the relative luminosity of the AGN and its host. 
Thus the near-IR is the best region in which to study the
host galaxy (McLeod \& Reike 1995). 
Host galaxy emission may also contribute to the
near- and mid-IR continuum (Figure~\ref{IRcolors}, Lumsden \& Alexander 2001)
and the optical continuum (Maiolino et al. 2000).

\subsection{Ingredients which determine the Shape of the SED}
Many parameters are important in determining the observed shape of
an AGN SED. The relative luminosities of the AGN itself,
related to the black hole mass, and of its host galaxy
determine the visibility of host galaxy emission. The accretion rate
and physical properties of the AD combine with its
inclination to our line-of-sight to determine the shape of the Big Blue
Bump. The amount, geometry, ionisation and optical depth of absorbing
dust and gas and its inclination determines the IR continuum and
the absorption of the optical, near-IR and soft X-ray continua.
The amount and location of scattering material and the
strength of the scattered light is also important in edge-on (type 2)
AGN where the primary AGN continuum is partially/fully obscured.
Finally the presence and strength of a radio source in the core of the AGN
affects the radio, far-IR and hard X-ray continua.


One property which, perhaps surpisingly, has not yet been shown to be
related to the SED is evolution. With the central black hole as driver of the
energy source, the black hole is expected to grow as
material is accreted and so the observable properties of the AGN
are expected to change as the source ages.
Although observations of high redshift AGN/quasars
are still limited, there is currently no convincing evidence for
a change in SED with redshift (Silverman et al. 2002,
Mathur et al. 2002, Brandt et al. 2002,
Vignali et al. 2003). Similarly, as discussed earlier, evolution
may unify ULIRGs and AGN predicting changes in the
IR SED as the dust first enshrouds and then heats up and disperses,
allowing the AGN to shine through. While the OUV and IR SEDs
show a wide diversity of shapes, there is no compelling evidence
for systematic evolution in their properties as a function of redshift.
The much larger number of high-redshift AGN now accessible
in many wavebands will provide significant new data on SED evolution.

\begin{figure}
\plotfiddle{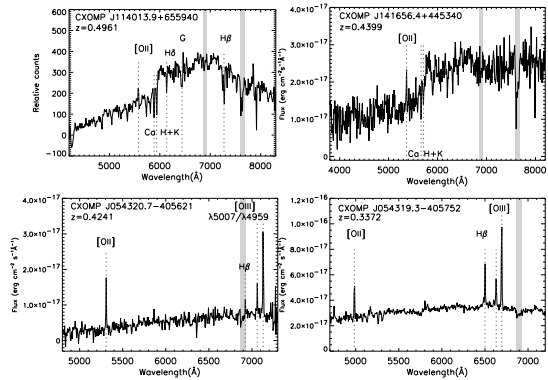}{3.5in}{0}{40}{40}{-120}{-20}
\vspace*{-1.7in}
\caption{Candidate buried AGN, ChaMP survey, Green et al. (2004).}
\label{buriedAGN}
\end{figure}

\section{Our Expanding View of AGN}

With so many variables, it is not surprising that AGN SEDs are so diverse,
it may be more surprising that we see such
similarities from source to source!
This leads to the question of selection. Are they mostly similar
because we select them to be? Would we recognise
an AGN that was, for example, so heavily absorbed that few if any of the
general characteristics we associate with an AGN are present?
Motivated by the large number of apparently normal galaxies being found by the
Chandra Deep Surveys, Moran et al. (2002) show how host
galaxy light may dominate optical AGN signatures in Sy2s
and thus hide a significant fraction of the
population.

Thus, perhaps the most critical open question is ``What
is an AGN?". New AGN are being found in radio
(Webster et al., 1995), IR (Cutri et al. 2001) and deeper optical surveys, 2DF
and SDSS (Richards et al. 2003). Models for the Cosmic X-ray
Background (CXRB) require a new population of X-ray absorbed AGN
(Gilli et al. 2001). Chandra and XMM-Newton
find X-ray sources sufficiently luminous to
be AGN but with no optical AGN characteristics
(Figure~\ref{buriedAGN}, Green et al. 2004 (ChaMP), 
Norman et al. 2002).
Similarly, bright
radio galaxies with strong, unresolved cores show no
optical AGN signatures. Are these all AGN? If so how do we relate them
to the more standard BEL AGN? What are a minimum set of
properties which define an AGN? How can we observe
the intrinsic population?

\begin{figure}
\plotfiddle{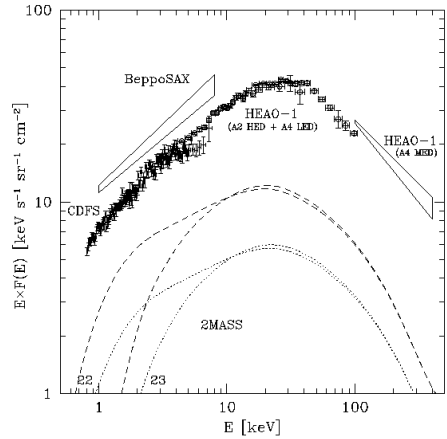}{3.0in}{0}{30}{30}{-100}{30}
\vspace*{-0.9in}
\caption{Contribution of 2MASS red AGN to the CXRB for two different
evolutionary models, Wilkes et al. 2003}
\label{2m_cxrb}
\end{figure}

Since AGN are multiwavelength emitters, we need a view in multiple
wavebands. For example the red AGN found in the 2MASS survey are 
hard, weak X-ray sources (Wilkes et al. 2002) and may contribute
$\sim 30$\% of the CXRB depending on their evolution 
(Figure~\ref{2m_cxrb}, Wilkes et al. 2003). But X-ray surveys must
cover more area to deeper flux levels to find the same kinds of sources.
The next generation of multi-wavelength surveys will provide both
deep, small areas (GOODS) and shallower, wider areas (SWIRE;
ChaMP, Silverman, this volume)
and sample the same/overlapping pieces of the intrinsic population.
The combination of sensitive IR and X-ray, provided by \chandra ,
XMM-Newton and SIRTF, is particularly powerful as the far-IR
sees all luminous IR sources while the X-rays select the
AGN from amongst the predominant IR galaxy population.
These are exciting times!

\end{document}